\documentclass[journal]{IEEEtran}
\usepackage{cite}
\ifCLASSINFOpdf
   \usepackage[pdftex]{graphicx}
\else
\fi
\usepackage{amsmath}
\usepackage{xcolor}

\usepackage{pdfpages}

\begin{document}

\null
\includepdf[pages=-]{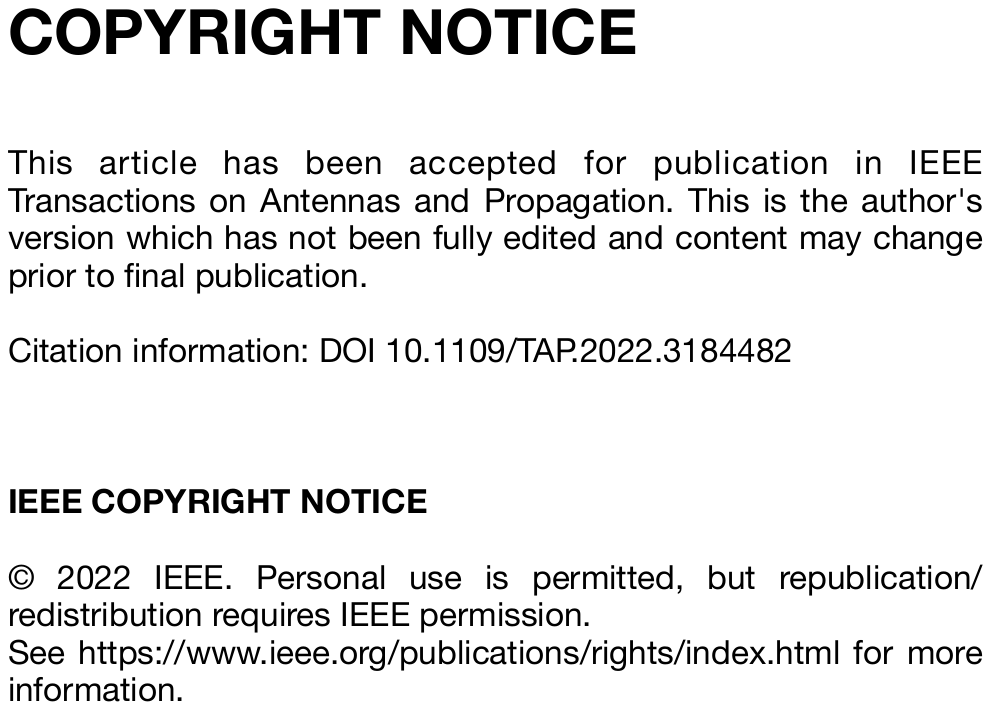}

\title{Tailoring Instantaneous Time Mirrors for\\Time Reversal Focusing in Absorbing Media}

\author{Crystal~T.~Wu,
        Nuno~M.~Nobre,
        Emmanuel~Fort,
        Graham~D.~Riley,
        and~Fumie~Costen,~\IEEEmembership{Senior~Member,~IEEE}% <-this % stops a space
\thanks{C. T. Wu is with the Department
of Computer Science, The University of Manchester, Manchester, M13 9PL,
UK (e-mail: crystal.wu@manchester.ac.uk).}% <-this % stops a space
\thanks{N. M. Nobre was with the Department
of Computer Science, The University of Manchester, Manchester, M13 9PL,
UK.}% <-this % stops a space
\thanks{E. Fort is with the Institut Langevin, ESPCI Paris, PSL University, CNRS UMR 7587, 1 rue Jussieu, Paris, 75005, France (e-mail: emmanuel.fort@espci.fr).}% <-this % stops a space
\thanks{G. D. Riley is with the Department
of Computer Science, The University of Manchester, Manchester, M13 9PL, UK (e-mail: graham.riley@manchester.ac.uk).}% <-this % stops a space
\thanks{F. Costen is with the Department
of Electrical and Electronic Engineering, The University of Manchester, Manchester, M13 9PL,
UK (e-mail: fumie.costen@manchester.ac.uk).}% <-this % stops a space
\thanks{Manuscript received July 7, 2021; revised April 8, 2022.}}



%%%\ifCLASSOPTIONpeerreview %%replace this with paper id number
\markboth{Journal of LaTeX Class Files,~Vol.~1, No.~1, July~2021}%
{Shell \MakeLowercase{\textit{et al.}}: Bare Demo of IEEEtran.cls for IEEE Journals}
%%%\fi%%replace this with paper id number



\maketitle

\begin{abstract}
The time reversal symmetry of the wave equation allows wave refocusing back at the source. However, this symmetry does not hold in lossy media. We present a new strategy to compensate wave amplitude losses due to attenuation. The strategy leverages the instantaneous time mirror (ITM) which generates reversed waves by a sudden disruption of the medium properties. We create a heterogeneous ITM whose disruption is unequal throughout the space to create waves of different amplitude. The time-reversed waves can then cope with different attenuation paths as typically seen in heterogeneous and lossy environments. We consider an environment with biological tissues and apply the strategy to a two-dimensional digital human phantom from the abdomen. A stronger disruption is introduced where the forward waves suffer a history of higher attenuation, with a weaker disruption elsewhere. Computer simulations show heterogeneous ITM is a promising technique to improve time reversal refocusing in heterogeneous, lossy, and dispersive spaces.
\end{abstract}


% \fi
\IEEEpeerreviewmaketitle

\section{Introduction} \label{sec:intro}

\IEEEPARstart{I}n the past few decades, the time reversal invariance of the wave equation has become a popular instrument to devise techniques for wave refocusing and scatterer localization. While radiation from a point source is characterized by a diverging forward propagation, the wave equation also admits a backward counterpart that can converge to the source location, now target. This behavior has been found to be reproducible through the time reversal mirror (TRM) and the instantaneous time mirror (ITM) which have recently been the subject of extensive numerical and practical experimentation~\cite{Wu:1992, Thomas:1996a, Prada:2002, Bond:2003, Larmat:2006, Olbricht:2013, Bacot:2016, Fink:2017}.

Studies have shown time-reversed waves can form foci accurately in a non-dissipative system, composed of lossless (but potentially heterogeneous) materials~\cite{Fink:1989, Fink:1992}. However, practical applications often involve lossy materials, in which the time reversal symmetry no longer holds due to attenuation~{\color{black}\cite{Tanter:1998, Yavuz:2005, Yavuz:2006}}. As a result, the obtained foci are usually of inferior quality. This applies for both TRM and ITM since they create reversed waves based on the same fundamental principle, the former using a spatial disruption and the latter a temporal disruption to the wave equation.

In this paper, we simulate ITM-induced reversed foci for electromagnetic waves traveling through a two-dimensional (2D) complex geometry with dispersive, i.e. frequency-dependent, biological tissues. We extend our previous work~\cite{Crystal:2020} and propose a new strategy based on exploratory experiments that refines the applied ITM by compensating wave attenuation. In the past, many methods have aimed to improve the quality of time-reversed foci, specifically in the context of TRM. They typically rely on the TRM transceivers to digitally manipulate the recorded signals between the stage of signal collection and re-emission{\color{black}~\cite{Yavuz:2005, Kosmas:2005, Yavuz:2006, Abduljabbar:2017, Mukherjee:2019}}. The strategy we developed is in contrast to TRM compensating techniques due to the absence of transceivers in ITM executions. 

Using numerical simulations, the feasibility of ITM in electromagnetic waves has been established through a \emph{time} disruption of the propagation speed via a sudden change in permittivity~\cite{Crystal:2020, Pacheco-Pena:2018}. Here, we exploit the relation between this disruption and its effect on the reversed wave amplitude. A heterogeneous ITM disruption, i.e. one which is unequal throughout space, is shown to successfully compensate unequally attenuated waves and to produce accurate foci.

We give an overview of our ITM implementation and a review of the elements that affect the reversed waves. The realization of our strategy is based on observations made from our studies in lossless materials. These ideas are then extended and applied to a 2D digital human phantom, a complex lossy environment. We show a case where a nearly accurate focus can be achieved without the proposed attenuation compensation strategy as well as two cases where compensation is paramount to improve the accuracy of the time-reversed foci, thereby supporting the benefits of our method.

{\color{black}This paper presents early research in a novel ITM compensating method which we have tested using numerical simulations. In practice, one can imagine an ITM could be achieved by rapidly modifying properties of the propagation medium such as humidity, pressure, and temperature~\cite{Kaatze:1989, Heidary:2010, Hermans:2014}. Our strategy, however, requires precise knowledge of the waves position in space at the moment of ITM, as well as the ability to alter the properties of the propagation medium with high spatio-temporal precision. Although these challenges are beyond the scope of this paper and remain a prospective area of research, the strategy may be first explored with other types of waves whose properties may be easier to manipulate, while research in techniques to manipulate the speed of light matures~\cite{Dutton:2004,Ginsberg:2007,Hau:2008}. For example, 2D waves, such as surface water waves~\cite{Apffel:2022}, allow an ITM to exploit a third dimension orthogonal to the planar area where propagation happens~\cite{Bacot:2017}.}

%\IEEEpubidadjcol

%%%%%%%%%%
%%%%%%%%%%
%%%%%%%%%%

\section{ITM for Electromagnetic Waves} \label{sec:itmtheory}

The concept of reversing waves by manipulating the time properties of a medium and without transceivers was first experimentally demonstrated by~\cite{Bacot:2016}. An ITM was introduced in water waves by implementing a disruption in celerity which subsequently generated time-reversed propagations visible to the naked eye. Our research in ITM-induced electromagnetic reversed waves is based on the same theoretical concept. We use a two-step ITM generation which, shortly after decreasing the propagation speed, brings it back to its initial value. {\color{black}This implementation ensures the wave speed never surpasses the speed of light in vacuum, no matter the medium.} In practice, the speed change entails a disruption of the wave impedance, on which the generation of ITM-induced electromagnetic reversed waves crucially depends~\cite{Morgenthaler:1958, Xiao:2014}.

As the phase velocity of electromagnetic waves is given by $v=1/\sqrt{\epsilon\mu}$, speed disruptions are possible through permittivity $\epsilon$ and/or permeability $\mu$ changes. We chose to keep $\mu$ constant and manipulate $\epsilon$, particularly the relative permittivity $\epsilon_\mathrm{r}$, as $\epsilon \equiv \epsilon(t) =\epsilon_{0}\epsilon_\mathrm{r}(t)$ where $\epsilon_{\mathrm{0}}$ is the vacuum permittivity. The phase velocity and wave impedance $\eta=\sqrt{\mu/\epsilon}$ are then time dependent. In general, we speak of time-varying media with time-varying propagating speed~\cite{Morgenthaler:1958, Fante:1971, Xiao:2014}.

{\color{black}
Time-reversed waves created by an ITM can be characterized in a manner similar to that of Fresnel's equations for light incident at a spatial boundary. Consider the first step of the temporal disruption. Assuming that the electric displacement field $\boldsymbol{D}$ and the magnetic flux density $\boldsymbol{B}$ are continuous at the time interface~\cite{Xiao:2014}, Fresnel's reflection and transmission coefficients at the time interface for the electric field $\boldsymbol{E}$ become~\cite{Morgenthaler:1958, Xiao:2014}:
\begin{align}
\label{eq:fresnelITMreflection}
    r \equiv &
    \frac{E_{\mathrm{r}}}{E_{\mathrm{in}}}
    =
    \frac{1}{2}
    \left( 
    \frac
    {\epsilon_{\mathrm{1}}}{\epsilon_{\mathrm{2}}}
    -\frac{\sqrt{\mu_{\mathrm{1}}\epsilon_{\mathrm{1}}}}{\sqrt{\mu_{\mathrm{2}}\epsilon_{\mathrm{2}}}}
    \right)
    =
    \frac{1}{2}
    \frac
    {\epsilon_{\mathrm{1}}}{\epsilon_{\mathrm{2}}}
    \left(1 
    -\frac{\eta_\mathrm{1}}{\eta_\mathrm{2}}
    \right), 
    \\
\label{eq:fresnelITMtransmission}
    t \equiv &
    \frac{E_{\mathrm{t}}}{E_{\mathrm{in}}} =
    \frac{1}{2}
    \left( 
    \frac
    {\epsilon_{\mathrm{1}}}{\epsilon_{\mathrm{2}}}
    +\frac{\sqrt{\mu_{\mathrm{1}}\epsilon_{\mathrm{1}}}}{\sqrt{\mu_{\mathrm{2}}\epsilon_{\mathrm{2}}}}
    \right)
    =
    \frac{1}{2}
    \frac
    {\epsilon_{\mathrm{1}}}{\epsilon_{\mathrm{2}}}
    \left(1 
    +\frac{\eta_\mathrm{1}}{\eta_\mathrm{2}}
    \right),
\end{align}
where subscripts 1 and 2 denote $\epsilon$, $\mu$, and $\eta$ prior to and after the disruption; and $E_{\mathrm{r}}$, $E_{\mathrm{t}}$, and $E_{\mathrm{in}}$ denote the electric field amplitudes of the reflected, transmitted, and incident waves, i.e. the time-reversed, forward, and initial forward waves, respectively. In the second step of the disruption, each of the two newly created waves then undergoes the reverse transition (exchange subscripts 1 and 2 above).}

\subsection{Amplitude Dependence on the Disruption Strength}
\label{subsec:ampcomp}

The sudden speed disruption can be modeled by a Dirac Delta function: ${v(t)^2 = v_0^2 [1 + \alpha\delta(t - t_\mathrm{ITM})]}$~\cite{Bacot:2016}. The phase velocity $v_0$ specifies the velocity prior to and following the ITM disruption instant $t_\mathrm{ITM}$ and $\alpha$ characterizes the intensity of the disruption. The reversed waves amplitude is proportional to $\alpha$, and therefore, to ${v^2/v_0^2 - 1 = \epsilon_{\mathrm{r}_0}/\epsilon_\mathrm{r} - 1}$ during ITM, where $\epsilon_{\mathrm{r}_0}$ is the relative permittivity corresponding to $v_0$. This relation is the primary object of our amplitude compensation strategy. Although attenuation is frequency-dependent, as $\alpha$ is set constant across all frequencies (i.e. $\epsilon_{\mathrm{r}_0}/\epsilon_\mathrm{r}$ is a constant), dispersive loss is not taken into account.

In a two-step ITM, the two speed changes create two replicas of the initial forward waves that are opposite in sign. If done in quick succession, the superimposed reversed waves approximate the derivative of the initial wave field, albeit flipped in sign. The sign flip is due to our implementation of ITM, which uses $\alpha < 0$. {\color{black}In a monochromatic wave, the derivative effect from an ITM can be described as a $\pi/2$ shift with respect to the initial wave if $\alpha > 0 $, or $-\pi/2$ if $\alpha < 0$.} The time interval between the two temporal disruptions {\color{black}which, in practice, is finite,} also has an effect on the amplitude of the reversed wave field. In our study, this time interval is kept consistent across all simulations. Additionally, $\epsilon_{\mathrm{r}}$ is increased by at most 100-fold, as it was experimentally observed that higher increments do not produce wave fields with significantly higher amplitude. In fact, this is to be expected since as $\epsilon_{\mathrm{r}}$ approaches $\infty$, $v$ tends towards 0 and $\alpha$ towards -1.

\subsection{Converging Waves and Diffraction Limit}
\label{subsec:switchsign}

Any converging wave that focuses continues its propagation as another diverging wave. Thus, a complete time reversal process also requires the source to be reversed, i.e. replaced by a sink to absorb the converging wave. We do not consider such sources. Without a time-reversed source, the re-diverging wave appears as a replica of the converging wave but opposite in sign~\cite{Cassereau:1992}. We demonstrate this visually in our ITM lossless media study in Section~\ref{sec:losslessstudy}.

An interesting phenomenon therefore arises from the interaction between the converging and re-diverging waves. As an example, one can picture this interaction by considering a converging wave shaped after the first derivative of the Gaussian function, where a trough is followed by a crest in space and time. After complete convergence from the trough, the re-diverging wave appears as a crest and reinforces the amplitude of the still converging crest. This interaction causes the spatial resolution of the focus point to be constrained by the diffraction limit, half a wavelength for a monochromatic wave~\cite{Cassereau:1992, Fink:2000, Rosny:2002}. In the case of a non-monochromatic wave, the limit depends on the frequency distribution of the reversed wave packet which, contrary to the monochromatic case, differs from that of the forward wave packet.

\subsection{Identifying an ITM Time-Reversed Focus}
\label{subsec:idmetrics}

We define the ITM-induced reversed focus time as ${t_{\mathrm{focus}} = 2t_{\mathrm{ITM}} - t_{\mathrm{src}}}$ where $t_{\mathrm{ITM}}$ is the speed disruption instant. Assuming a causal excitation function of finite support~\cite{Cassereau:1992}, $t_{\mathrm{src}}$ is an arbitrary instant of choice during source excitation (e.g. the mid instant for a symmetric excitation). For a two-step ITM of non-zero duration, $t_\mathrm{ITM}$ is defined as the mid instant between the two disruptions. This metric assumes the wave speed is the same regardless of its traveling direction in the same medium. This, however, only holds in non-dispersive media. In practice, although both ITM and absorption alter the field's frequency spectrum and the assumption might break, we found $t_{\mathrm{focus}}$ to be a good approximation.

The focus identified at $t_{\mathrm{focus}}$ in lossy and heterogeneous media poses another question. The path traced by a diverging initial wave might not be equally absorbing for every direction relative to the source position. This heterogeneity causes path dependent attenuation and deforms the wave. As a result, the location which the reversed wave converges to may be shifted away from the source and deemed inaccurate. To characterize the properties of the identified focus at $t_{\mathrm{focus}}$, we use the quantitative metrics in Section~\ref{subsec:analyticalmetrics}.

Finally, the properties of the time-reversed focus at $t_{\mathrm{focus}}$ depend on the source excitation function and the selected $t_{\mathrm{src}}$, even in vacuum or any other lossless environment. For a non-time-reversed source in a 2D geometry, an excitation of even-type symmetry about $t=t_{\mathrm{src}}$ would, in theory, cause the field to \emph{vanish everywhere} at $t_{\mathrm{focus}}$. An excitation of odd-type symmetry about $t=t_{\mathrm{src}}$ where the amplitude is zero, i.e. the point $(t_{\mathrm{src}},0)$, would cause the field to \emph{peak at the target location} at $t_{\mathrm{focus}}$. In this work, we use excitations of the latter sort to simplify the analysis, and assume this when defining the metrics in Section~\ref{subsec:analyticalmetrics}.

\subsection{Analyzing an ITM Time-Reversed Focus}
\label{subsec:analyticalmetrics}

A time-reversed focus is assessed on four quantitative features, all evaluated at or around $t_{\mathrm{focus}}$. The first two are the \emph{spatial accuracy} $A$ and \emph{spatial resolution} $R$ at $t_{\mathrm{focus}}$. The accuracy $A$ is the distance between the field's maximum amplitude in space and the known source location (i.e. the intended target location), whereas the resolution $R$ is the contour area at 50\% amplitude around the same maximum, capturing the 2D full width at half maximum~\cite{Crystal:2020}.

Third, to characterize the evolution of the converging focus, we introduce the \emph{time persistence} $P$, to quantify for how long the field energy ``lingers'' around the source location:
\begin{equation}
\label{eq:timepersist}
P = \frac{1}{|\Delta t_\mathrm{focus}|} \sum_{t \in \Delta t_\mathrm{focus}} [\mathrm{A}(t) \leq \lambda_{\mathrm{rev}}/8],
\end{equation}
where we have used the Iverson bracket notation. The accuracy $A(t)$ is dependent on time $t$, $\lambda_{\mathrm{rev}}$ is the central wavelength of the reversed wave in vacuum (irrespective of the propagation media), and $\Delta t_\mathrm{focus}$ is a (discretized) time window around $t_\mathrm{focus}$. We define
\begin{equation*}
\Delta t_\mathrm{focus} = \left\{ \left \lceil{t_\mathrm{focus}-T_\mathrm{rev}/4}\right \rceil   ,~\dots ~,\left \lfloor{t_\mathrm{focus}+T_\mathrm{rev}/4}\right \rfloor \right\},
\end{equation*}
where $T_\mathrm{rev}$ is the time period of $\lambda_{\mathrm{rev}}$.
%%%%%%%%%%%%%%%%%%%%%%%%%%%%%%
Empirical analysis shows time persistence is proportional to the wavelength and inversely proportional to the wave speed, suggesting that $P$ is an invariant in the absence of absorption. The fractions of $T_\mathrm{rev}$ and $\lambda_{\mathrm{rev}}$ defined were chosen to allow relative comparisons of the different foci in this paper, however, other choices can be made as long as the comparisons remain meaningful.

Fourth, we utilize \emph{{\color{black}minimum entropy deconvolution}}\footnote{{\color{black}To avoid ambiguity, the minimum entropy deconvolution, introduced by Ralph A. Wiggins, is not related to the thermodynamic entropy~\cite{Wehrl:1978} or the Shannon entropy~\cite{Shannon:1948}.}}{\color{black}~\cite{Wiggins:1978, Wu:1998}}{\color{black}, also known as simply {\em entropy} in the time reversal field~\cite{Kosmas:2005, Mukherjee:2019},} to characterize the ``concentration'' of the field at $t_\mathrm{focus}$ relative to other foci at different times. This method has been used to identify the refocusing instant without tracking the field in any particular location~\cite{Kosmas:2005, Mukherjee:2019}. It assumes a time-reversed focus in space would exhibit a localized field with minimal disorder everywhere else. For a discretized 2D space grid, the entropy $S$ can be expressed by {\color{black}the inverse varimax norm}

\begin{equation}
  \label{eq:entropy}
S(t) = \frac{ \bigg[ \sum\limits_{x,y}  u_\mathrm{E}(x,y,t) \bigg] ^{2}}
{\sum\limits_{x,y} u^2_\mathrm{E}(x,y,t)}, 
\end{equation}
where the density of the energy stored in the electric field is

\begin{equation}
  \label{eq:energydensity}
    u_\mathrm{E}(x,y,t) = \frac{1}{2} E_z(x, y,t)D_z(x, y,t)
\end{equation}
since the simulations use the transverse magnetic mode of the finite difference time domain method. 

However, a local minimum in $S$, even if around $t_\mathrm{focus}$, does not guarantee an accurate focus. Moreover, in Section~\ref{sec:losslessstudy} we show that an accurate focus is possible without a local entropy minimum. Hence, we solely utilize entropy plots to describe how localized a focus is when compared to the field distribution at other times.

%%%%%%%%%%
%%%%%%%%%%
%%%%%%%%%%

\section{ITM in Lossless Materials}
\label{sec:losslessstudy}

We present experiments in air, a lossless medium, that motivate our proposed strategy to improve focus quality in complex lossy spaces such as the human body.

Our previous work~\cite{Crystal:2020} showed successful creation of time-reversed waves by applying a homogeneous ITM to the entire space of a homogeneous medium. The results are shown to be similar to those of TRM, given a simple geometry. In this section, we perform ITM only in selected regions of the space and use the metrics described in Section~\ref{subsec:analyticalmetrics} to analyze the conducted simulations. The objective is to mimic regions with nonexistent or very small amplitude reversed waves traveling towards the source, as may happen in a complex environment due to absorption. The lessons learned are then extrapolated to develop attenuation compensation techniques.

\subsection{Simulation Model and Configuration}
\label{subsec:simLossless}

The space, homogeneously filled with air, was 1000~mm $\times$ 1000~mm in size with absorbing boundary conditions~\cite{Wrenger:2002}. A soft point source~\cite{Costen:2009} was placed in the center and modeled by an excitation current using the first derivative of a Gaussian pulse covering a frequency range from approximately 0.1~GHz to 7.5~GHz, centered at 3.0~GHz. We used the finite difference time domain method and set the time step and the length of a cell to 1.92~ps and 1~mm, respectively.

To induce ITM, a rapid interval of two speed disruptions was introduced in the selected regions, where air's $\epsilon_{\mathrm{r}}$ was increased linearly by 100-fold, briefly kept constant, then linearly returned to its initial value over three periods of approximately 10 time steps. Thus, we create a time-reversed wave which differentiates (and flips in sign, see Section~\ref{subsec:ampcomp}) the initial wave along the radial coordinate (since there is circular symmetry), as predicted and observed in~\cite{Bacot:2016, Bacot:2017}.

\subsection{Lessons from Heterogeneous ITM}
\label{secsec:losslesslesson}

\begin{figure}
    \centering
    \includegraphics[width=\columnwidth]{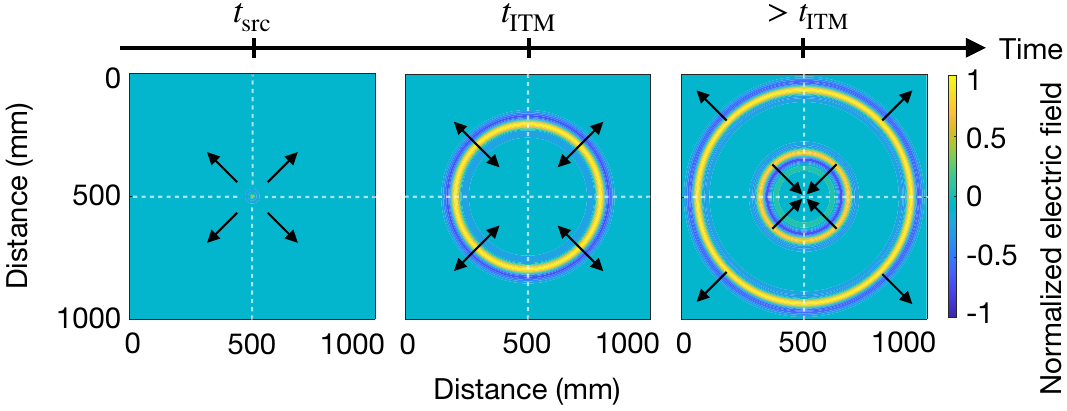}
    \caption{Evolution of the normalized electric field for waves induced by a point source which were then subjected to a homogeneous ITM applied to the entire simulation space. The space is filled with air homogeneously and excited at the point where the two white dashed lines meet.}
    \label{fig:baseline}
\end{figure}

Fig.~\ref{fig:baseline} demonstrates a homogeneous ITM (disruption in the entire space). Also shown are the earlier time $t_\mathrm{src}$ and a snapshot of a later time depicting both the forward (diverging) and the backward (converging) wave.

\begin{figure*}
    \centering
    \includegraphics[width=2\columnwidth]{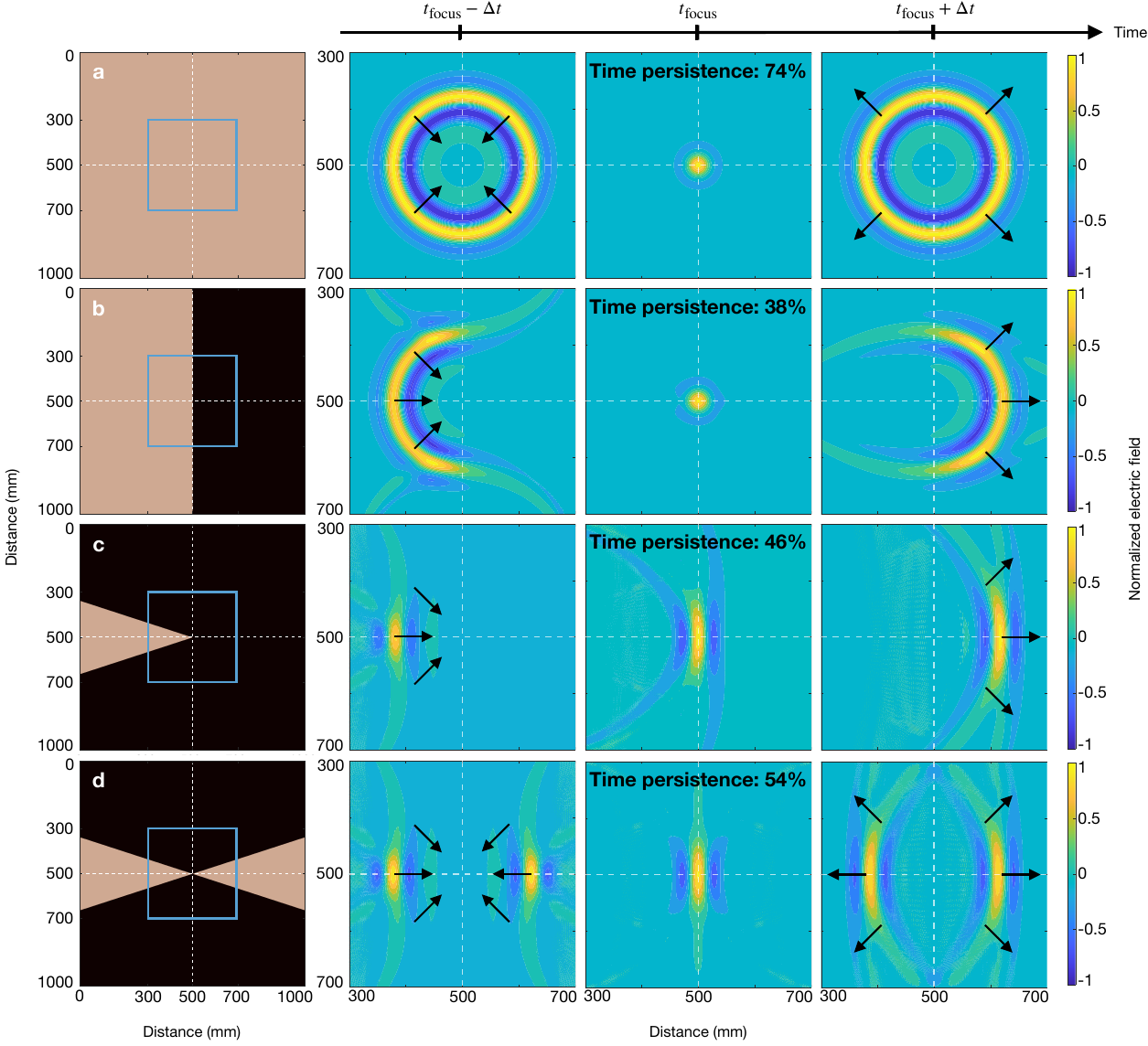}
    \caption{ITM in selected portions of a homogeneous lossless space, air, and the resulting time-reversed foci. (a) Homogeneous ITM, continues the timeline shown in Fig.~\ref{fig:baseline}. (b), (c), and (d) Heterogeneous ITM in different scenarios. On the left, ITM was applied to the brown regions while the properties in the dark regions remained the same. On the right, time sequence of the normalized electric field distributions are shown corresponding to the scenarios depicted to their left. The target location is marked by where the two white dashed lines meet. We show the time persistence at $t_\mathrm{focus}$ on the top of the corresponding snapshot. The field distribution is given for the region within the blue box.}
    \label{fig:simple}
\end{figure*}

\begin{figure}
    \centering
    \includegraphics[width=\columnwidth]{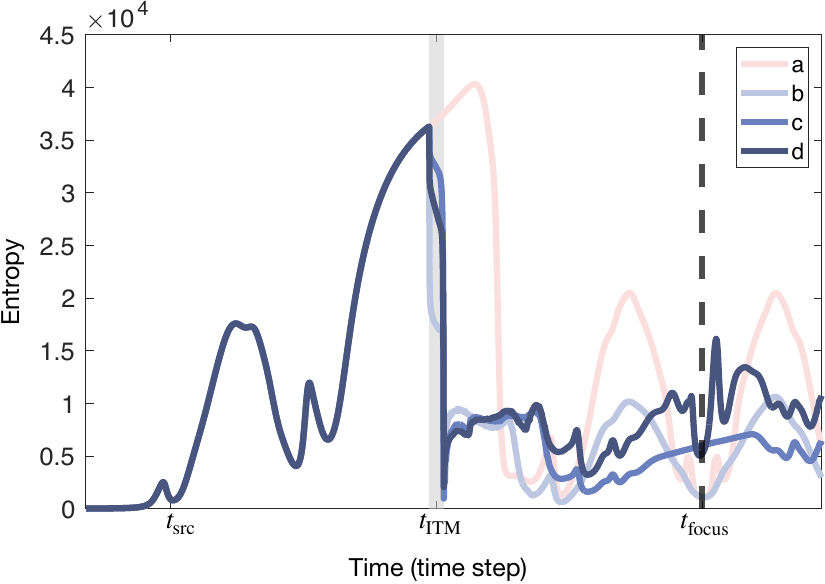}
    \caption{Entropy in the space within the blue box shown in Fig.~\ref{fig:simple}.  (a) Homogeneous ITM{\color{black}, where the temporal disruption is performed everywhere in space.} (b), (c), and (d) Heterogeneous ITM {\color{black}cases, where the temporal disruption is performed non-uniformly in space, as depicted by the brown and dark regions on the left} in Fig.~\ref{fig:simple}. The grey area shows the ITM interval.}
    \label{fig:simpleR}
\end{figure}

We performed ITM as illustrated on the left of Fig.~\ref{fig:simple}. With the exception of part (a), where ITM is performed homogeneously, parts (b) to (d) show heterogeneous ITM experiments. The $\epsilon_{\mathrm{r}}$ is manipulated solely in the brown region of the space. Air's properties in the dark regions remain unchanged so no reversed waves are generated in those regions. The figure also shows the resulting normalized electric field distribution within a 400~mm $\times$ 400~mm box with its center affixed to the center of the simulation space, i.e. the source location. We show a time sequence including the predicted convergence moment $t_\mathrm{focus}$ and instants $\Delta t = 200$ time steps to its past and future. Notice how in Fig.~\ref{fig:simple}, the converging and re-diverging waves appear to be mirrored: the four peaks switch their signs before and after $t_{\mathrm{focus}}$ as they continue to propagate.

Fig.~\ref{fig:simpleR} shows the entropy in (\ref{eq:entropy}) for all four scenarios within the defined box around the source. As seen in Figs.~\ref{fig:baseline} and \ref{fig:simple}, the defined box excludes the forward wave around $t_\mathrm{focus}$ and exposes features from the reversed field. The circular symmetry of case (a) determines a local entropy minimum at $t_{\mathrm{focus}}$, and we found the focus resolution is of the order of half a central wavelength of the reversed wave. Case (b) also exhibits an entropy minimum around $t_{\mathrm{focus}}$. Its time sequence in Fig.~\ref{fig:simple}b shows the focus keeps a circular shape at $t_{\mathrm{focus}}$ and is remarkably similar to the outcome obtained in case (a). However, the lack of a converging reversed wave field from half of the space incurs a drop in time persistence from 74\% in case (a) to 38\% in case (b).

In case (c), the change in $\epsilon_{\mathrm{r}}$ is only applied to the brown area sweeping a narrow polar angle range. Fig.~\ref{fig:simpleR}c shows the entropy has no local minimum around $t_{\mathrm{focus}}$, suggesting a focus has failed to form. However, a closer look at Fig.~\ref{fig:simple}c reveals a focus successfully forms, albeit less localized due to the partial reversed wave. The takeaway, as suggested at the end of Section~\ref{subsec:analyticalmetrics}, is the minimum entropy deconvolution alone might not be a reliable method to identify a focus.

In case (d), we mirrored the area in case (c) and applied ITM to the two opposing brown areas. Its entropy in Fig.~\ref{fig:simpleR}d shows a local minimum around $t_{\mathrm{focus}}$.
The distinctness of case (d) when compared to cases (b) and (c) stems from its higher time persistence at 54\%, underlining the importance of modifying ITM to ``balance'' the reversed wave along opposing directions. 

%%%%%%%%%%
%%%%%%%%%%
%%%%%%%%%%

\section{Bespoke Heterogeneous ITM}
\label{sec:refineditm}

Although time reversal is capable of recovering reflected, refracted, and scattered waves~\cite{Fink:1992, Prada:2002}, there are two important shortcomings. First, it is well known that reduced wave intensities due to attenuation in lossy media are not recoverable by the time reversal process. Second, in the case where it is \emph{impossible} 
to implement ITM in the whole space, e.g. the region where the source resides is deemed inaccessible, the intensities of the reversed waves are further reduced, even if the space is lossless. As a consequence, reversed waves in lossy media or generated by an incomplete ITM might not be able to retrace their paths back to the source.

\begin{figure}
    \centering
    \includegraphics[width=\columnwidth]{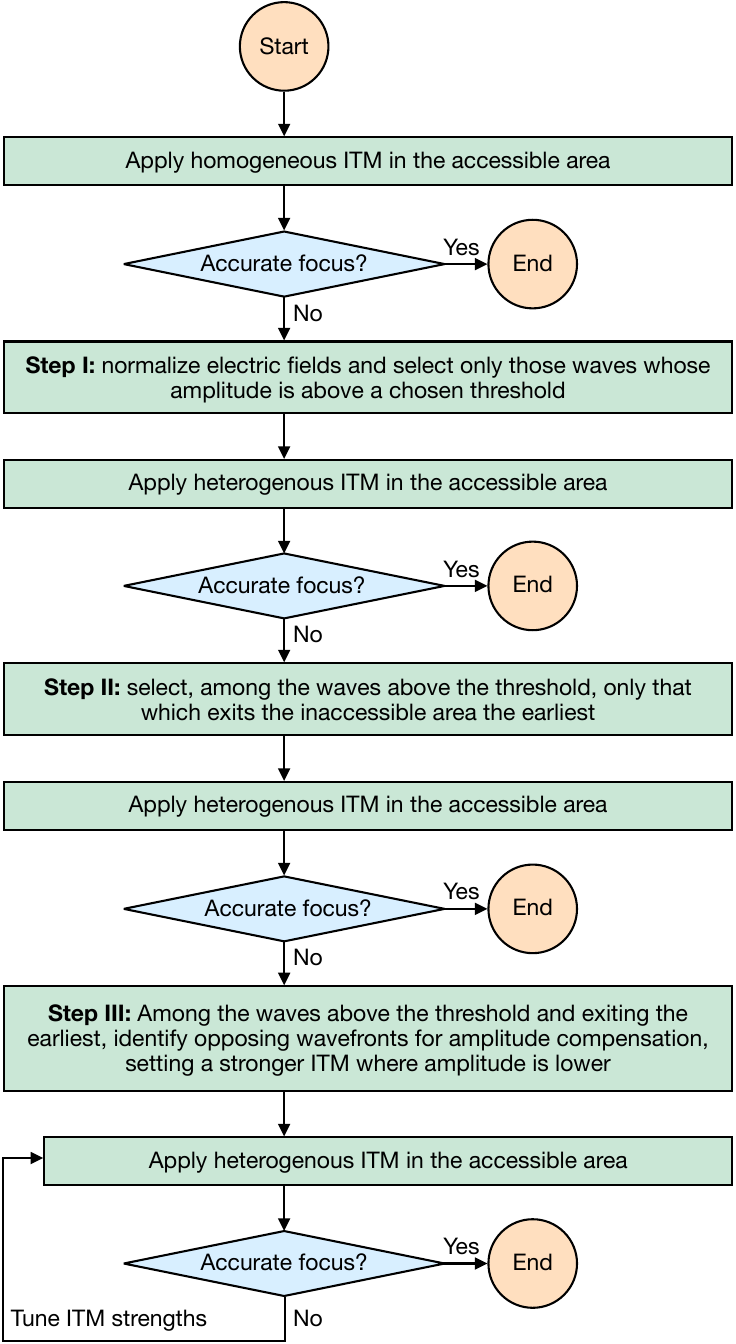}
    \caption{{\color{black}A summary of the heterogeneous ITM strategy.}}
    \label{fig:blockdiagram}
\end{figure}

The observations made in the lossless study were used to guide a new strategy that is able to improve the localization of a time-reversed focus in complex environments. The idea is to not only manipulate where a reversed wave is generated but also to adjust its amplitude. {\color{black} The proposed strategy is presented as a flowchart in Fig.~\ref{fig:blockdiagram}.} As digital conditioning is unavailable in ITM due to the lack of transceivers, our strategy instead exploits the \emph{effect localized $\epsilon_{\mathrm{r}}$ changes have on the generated reversed wave amplitude and thus geometry}. 

\subsection*{{\color{black}Step I: Reducing} Destructive Interference}
\label{subsec:avoideDestInterf}

We first normalize the electric field {\color{black}from the entire simulation space} according to the absolute maximum in space at the first speed disruption.
We then apply a threshold to filter the initial forward waves, such that only those whose normalized amplitude is \emph{above the threshold} undergo the ITM disruption. The threshold is chosen to have a near zero value to avoid introducing large discontinuities in the wave field{\color{black}, which the differentiating nature of ITM could aggravate. This step essentially eliminates one of the field signs, i.e. either all of the troughs or all of the crests, from the ITM in an attempt to minimize the destructive interference arising from peaks of opposite sign. In case an inaccurate focus remains, the initial waves above the threshold then proceed to Step II.}

\subsection*{{\color{black}Step II: Utilizing the Least Attenuated Initial Waves}}
\label{subsec:useLeastAttenuated}

The aim of this step is to perform ITM only in a selected part of the (filtered) initial wavefronts {\color{black}that have experienced attenuation the least.} As propagation speeds and absorption characteristics differ depending on the media the waves traverse, the wave may be “fragmented” or “splintered” into several pieces. This step considers only the wave that reaches the ITM accessible region the earliest, which is then marked for ITM. {\color{black}The combined effect of Steps I and II prepares} the generation of ITM reversed waves based {\em solely} on a single peak{\color{black}, a crest or a trough,} of the initial {\color{black}wavefront which has undergone minimum attenuation.} Thus, the number of crests and troughs in the reversed wave are reduced to precisely one crest and one trough (recall ITM differentiates the field) regardless of the number of such displacements in the original source excitation, which helps preventing the field maximum at $t_{\mathrm{focus}}$ to shift away from the target. Waves below the threshold in Step I {\color{black}and those that are unmarked for ITM in Step II will simply} carry on and diverge as initial forward waves{\color{black}, without reversed wave generation.}

\subsection*{{\color{black}Step III:} Directional Information for Amplitude Compensation}
\label{subsec:gatherdirampcomp}

{\color{black}To compensate for the imbalanced amplitude caused by path-dependent attenuation,} {\color{black}in addition to Step II}, we identify those initial wavefronts that experienced attenuation the least for no less than two opposing directions. {\color{black}We group a number of initial wavefronts according to their exit site from the inaccessible area,} {\color{black} pair these groups along, approximately, opposing directions and} track their propagation path until the ITM instant. {\color{black} When the ITM is performed, the aim is to, for each pair,} create higher intensity reversed waves in the region that is associated with the group of initial waves of lower amplitude, and to create lower intensity reversed waves in the region which is associated with the group of initial waves of a higher amplitude. {\color{black}In essence, Step III tries to form a focus with contributions from reversed waves with a geometry similar to that of part (d) in Fig.~\ref{fig:simple}, rather than that of part (c).}

%%%%%%%%%%
%%%%%%%%%%
%%%%%%%%%%

\section{Heterogeneous ITM in Lossy Materials}
\label{sec:lossystudy}

This section applies the proposed strategy to ITM in
a large lossless space surrounding an inaccessible complex lossy region, namely a human phantom in air whose biological tissues act as a low pass filter. In each scenario, homogeneous ITM is simulated as a baseline where the $\epsilon_{\mathrm{r}}$ change is uniform in the lossless space, as in Fig.~\ref{fig:simple}a. To analyze the proposed strategy, heterogeneous ITM is performed, i.e. the change in $\epsilon_{\mathrm{r}}$ is applied to only some parts of the lossless space, similar to parts (b) to (d) of Fig.~\ref{fig:simple}. Depending on the scenario, we apply heterogeneous ITM with or without the amplitude compensation technique described in Section~\ref{subsec:gatherdirampcomp}, where the $\epsilon_{\mathrm{r}}$ change varies across the selected waves in the lossless space.

\subsection{Simulation Model and Configuration}
\label{subsec:simLossy}
The simulations were carried out using the frequency-dependent finite difference time domain method with a one-pole Debye relaxation model for $\epsilon_{\mathrm{r}}$~\cite{Costen:2009}:
\begin{equation}
  \label{eq:onePoleDebye}
  \epsilon_{\mathrm{r}} = \epsilon_{\mathrm{\infty}} + 
  \frac{\epsilon_{\mathrm{s}} - \epsilon_{\mathrm{\infty}}}{1 + i\omega\tau_{\mathrm{D}}} - i\frac{\sigma}{\omega\epsilon_{\mathrm{0}}},
\end{equation}
where $\epsilon_{\mathrm{\infty}}$ is the optical relative permittivity, $\epsilon_{\mathrm{s}}$ the static relative permittivity, $i$ the imaginary unit, $\omega$ the angular frequency, $\tau_{\mathrm{D}}$ the relaxation time, and $\sigma$ the conductivity. Table~\ref{tab:debyeParameters} shows the materials used in our simulation and their corresponding Debye parameters~\cite{Abalenkovs:2011, Hemmi:2014}. As the environment changes with time in ITM, so do its Debye parameters~\cite{Costen:2005}. During ITM, air's properties momentarily change as explained in Section~\ref{sec:itmtheory}. All biological tissues keep their properties throughout time.

\begin{figure}
    \centering
    \includegraphics[width=.8\columnwidth]{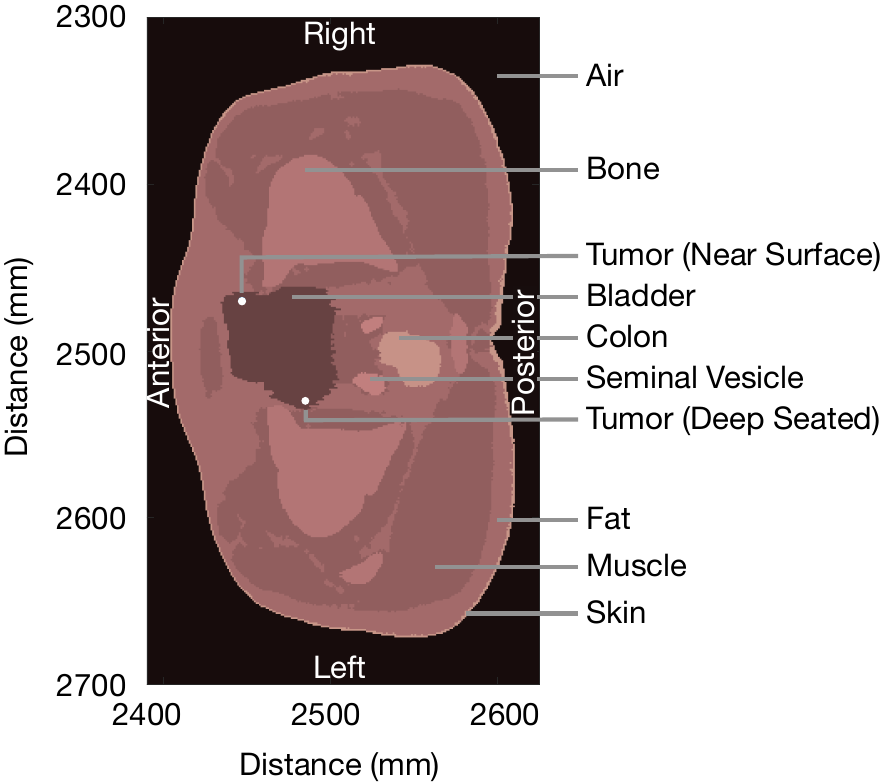}
    \caption{2D digital human phantom from an axial section of an adult male's lower abdomen. The near surface and the deep seated tumors are centered at (2447, 2470) and (2485, 2530), respectively.}
    \label{fig:phantom}
\end{figure}

Since the space contains lossy media and the source resides in such a region, some reflected and refracted waves remain near the source instead of propagating outwards into air. These waves are not an issue in TRM as reversed waves are sent into a space with no fields. In ITM however, reversed waves are created as soon as the speed disruption happens. Thus, to facilitate localization and wave focusing at the target with minimal ``leftover'' noise, we induce ITM at a sufficiently late time to allow the intensity of the leftover waves to decline considerably below the converging reversed waves, thereby minimizing interference. Since the initial forward waves continue to propagate away from the source, this implies modeling a sufficiently large space. We enlarged the simulation space to 5000~mm~$\times$~5000~mm and placed, at the center, a 2D axial section of a digital human phantom from the lower abdomen (1~mm resolution). Fig.~\ref{fig:phantom} shows its geometry.

\begin{table}
  \caption{Debye parameters of the materials used in the simulations.}
  \label{tab:debyeParameters}
  \begin{center}
    \begin{tabular}{lrrrr}
      \hline
      Material & $\epsilon_{\mathrm{s}}$ & $\epsilon_{\mathrm{\infty}}$ & $\sigma$~(S/m) & $\tau_{\mathrm{D}}$~(ps)\\
      \hline
    
      Air              & ~1.00 & 1.00  & ~0.00 & 0.00\\
      Bladder          & 19.33 & ~9.67 & 0.30  & 20.84\\
      Bone             & 14.17 & ~7.36 & 0.10  & 34.11\\
      Colon            & 61.12 & 34.67 & 0.72  & 31.17\\
      Fat              & ~5.53 & ~4.00 & 0.04  & 23.63\\
      Muscle           & 56.93 & 28.00 & 0.75  & 18.67\\      
      Seminal Vesicle  & 60.54 & 27.71 & ~0.81 & 17.71\\
      Skin             & 47.93 & 29.85 & 0.54  & 43.63\\
      
      \rule{0pt}{3ex}Tumor & 20.26 & 10.64 & 0.74 & 20.84 \\
      \hline
    \end{tabular}
  \end{center}
\end{table}

Two sets of experiments were carried out: the first uses a deep seated source, and the second a near surface source, both using the same settings as Section~\ref{subsec:simLossless}.
The goal is to utilize reversed waves to refocus at the targeted source location. The point source was placed within an artificially introduced tumor-like scatterer of 5~mm diameter inside the bladder with the Debye parameters in Table~\ref{tab:debyeParameters}. These Debye parameter values represent a conservative increase of approximately 10\% in the complex relative permittivity relative to healthy bladder tissue over the frequency range of interest. This choice is based on the general consensus that there is a significant difference between healthy and malignant bladder tissues, though the literature disagrees on the specific characteristics~\cite{Ibrahim:2012, Porter:2018, Fahmy:2020}.

\subsection{A Deep Seated Source}
\label{subsec:deepseated}

\begin{figure}[!b]
    \centering
    \includegraphics[width=\columnwidth]{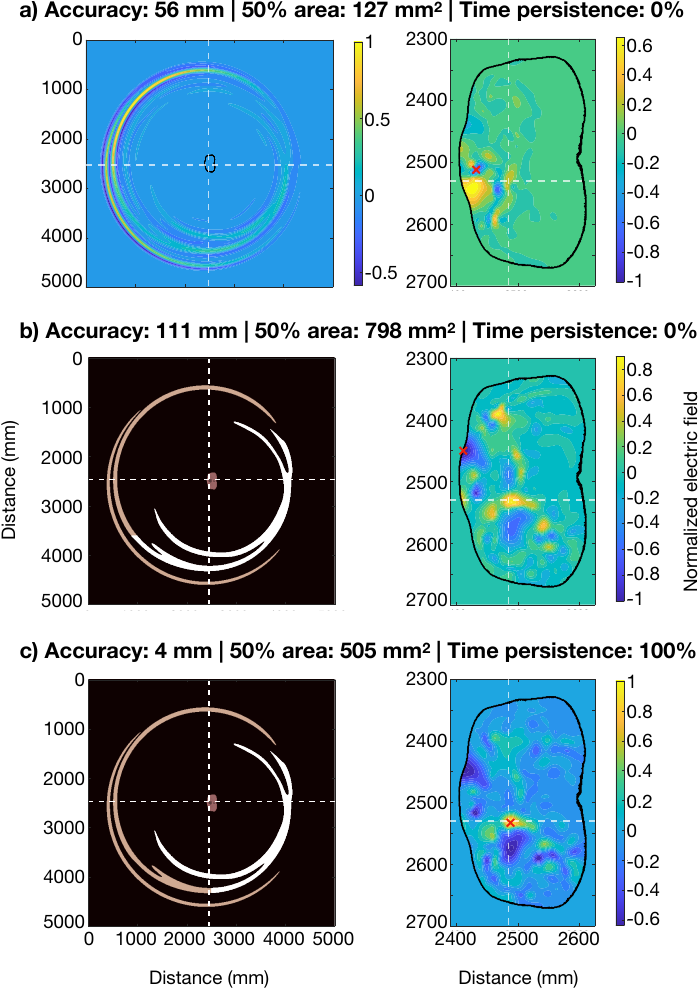}
    \caption{{\color{black}Deep seated scenario.} (a) {\color{black}Control:} homogeneous ITM. (b) and (c) {\color{black} Case study:} heterogeneous ITM {\color{black} with all three steps of the strategy applied,} including amplitude compensation.  {\color{black}On the left, illustration of} (a) the initial wave field at the first disruption instant, and (b) and (c) two scenarios of the initial wavefronts selected for an ITM of two strengths {\color{black}along opposing directions.} The results at $t_{\mathrm{focus}}$ are shown on the right.}
    \label{fig:deep}
\end{figure}

We first review a deep seated source placed within the bladder tumor shown in Fig.~\ref{fig:phantom}. A homogeneous ITM of the same nature used in the lossless study part (a) is applied to the entire space of air surrounding the biological tissues. This is the control case. The normalized electric field at the instant of the first speed disruption is shown in Fig.~\ref{fig:deep}a (left). 

\begin{figure}[!b]
    \centering
    \includegraphics[width=\columnwidth]{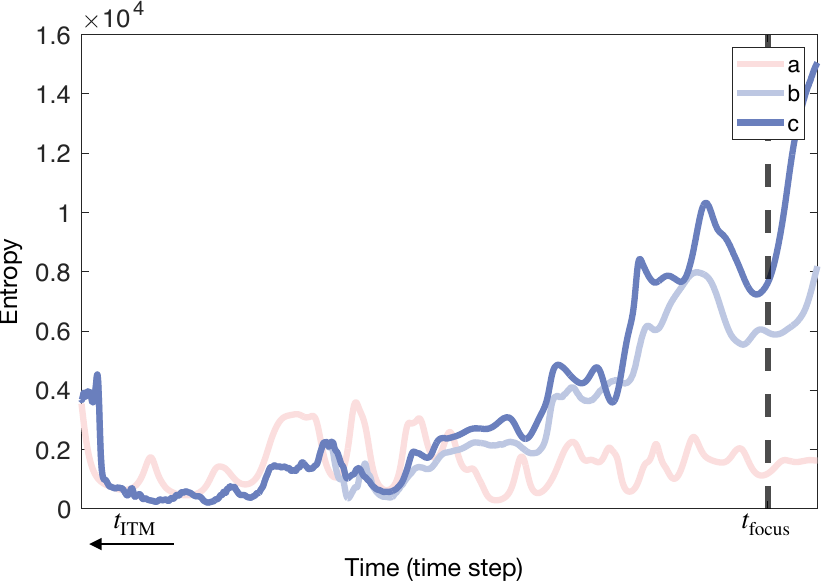}
    \caption{Entropy {\color{black}within the digital human phantom} of the deep seated scenario {\color{black}depicted in Fig.~\ref{fig:deep}.} (a) {\color{black}Control:} homogeneous ITM. (b) and (c) {\color{black}Case study:} heterogeneous ITM {\color{black}with all three steps of the strategy applied, including amplitude compensation. The temporal disruption is performed non-uniformly in air, as depicted by the white, brown, and dark regions in Fig.~\ref{fig:deep}.}}
    \label{fig:deepR}
\end{figure}

The control case is compared with two experimental cases to which we applied the proposed heterogeneous ITM strategy: filtering with a 5\% threshold in Step I, selecting the least attenuated wavefront in Step II, and compensating attenuation in Step III. Observe the wave field is affected by the source position and the tissue composition, particularly the large posterior muscle mass. Posterior waves are therefore overall lower in amplitude and closer to the source than anterior waves, providing a hint regarding the two opposing directions into which the wavefronts should be grouped.

The selected wavefronts are shown in Fig.~\ref{fig:deep}b (left), and they are grouped with respect to where they exited the inaccessible area, as discussed in Section~\ref{sec:refineditm}, Step III. This guarantees the lower amplitude ``tails'' extending to the posterior side are grouped with the higher amplitude wavefront exiting from the anterior side. However, this also poses a problem on the overlapping region just outside the left-hand side of the body.

Since waves exiting the digital human phantom from different locations overlap as they diverge in air, Fig.~\ref{fig:deep}c (left) shows a second grouping alternative. The difference between parts (b) and (c) of the figure is in the highlighted brown and white regions. Consideration of the time evolution of the initial wave suggests that some of the white segment in Fig.~\ref{fig:deep}b (left) is part of the anterior wavefront. As the area is ambiguous and it is unclear to which wavefront the tails belong, we experimented with the amplitude compensation using both partitioning choices.

Experimental observation determined that a change of 100-fold in $\epsilon_{\mathrm{r}}$ for the white wavefront and of 1.02-fold for the brown wavefront produces favorable results. We show their normalized electric field distributions within the digital human phantom at $t_{\mathrm{focus}}$ on the right of Fig.~\ref{fig:deep}, corresponding to the setups to their left. In this paper, we present the refocusing qualities for each experiment on the top of the respective figure. At $t_{\mathrm{focus}}$, the field maximum within the phantom is denoted by a red cross and the fields in air are masked to zero.  The target lies on the intersection of the white dashed lines and the skin of the human body is outlined in black.

The results for the control case in Fig.~\ref{fig:deep}a show a degraded time reversal focus due to attenuation, with the field predominantly locating near the anterior border of the digital human phantom. Although a peak is formed around the source, the maximum peak is 56~mm away from the targeted site. From experimental cases (b) and (c), results of the electric field distribution at $t_\mathrm{focus}$ show clear improvements in terms of focusing localization at the target. In Fig.~\ref{fig:deep}b, the field amplitude near the skin of the digital human phantom at (2400, 2450) is seen much stronger than that at the targeted site. On the other hand, case (c) shows a maximum peak with 4~mm accuracy, 100\% time persistence, and a 505~mm$\mathrm{^2}$ contour area at half the maximum amplitude. The 100\% time persistence in case (c) is likely a byproduct of absorption in the biological tissues, which attenuates the higher frequency components.

Fig.~\ref{fig:deepR} shows the entropy of the control and experimental cases for the deep seated source scenario, starting around the time when waves begin to enter the human body. For all cases, a local minimum exists near $t_{\mathrm{focus}}$. However, the entropy of the experimental cases has an upward trend as time progresses towards and past $t_{\mathrm{focus}}$ while it remains somewhat constant in the control case. Clearly, the local minimum in case (b) does not indicate an accurate focus. However, the strong local minimum in case (c) supports the fact that the obtained time-reversed focus is highly concentrated as seen in Fig.~\ref{fig:deep}c.

\subsection{A Near Surface Source}

\begin{figure}[!b]
    \centering
    \includegraphics[width=\columnwidth]{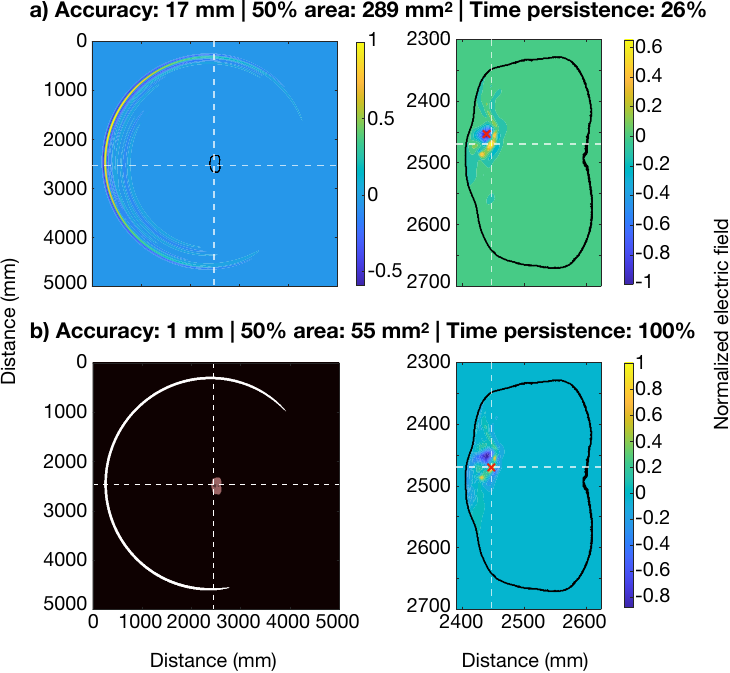}
    \caption{{\color{black}Near surface scenario.} (a) {\color{black}Control:} homogeneous ITM. (b) {\color{black}Case study: }heterogeneous ITM {\color{black}with Steps I and II of the strategy applied}. On the left, {\color{black}illustration of} (a) the initial wave field at the first disruption instant, and (b) the wavefront selected for ITM. The results at $t_{\mathrm{focus}}$ are shown on the right.}
    \label{fig:near}
\end{figure}

We review a scenario where the source is still within the bladder but closer to the physical interface between the human body and air. The position of the near surface point source (and tumor) is illustrated in Fig.~\ref{fig:phantom}. Akin to the deep seated source scenario, a control case is generated using a homogeneous ITM. The normalized electric field at the first instant of ITM disruption is shown in Fig.~\ref{fig:near}a (left).

\begin{figure}[!b]
    \centering
    \includegraphics[width=\columnwidth]{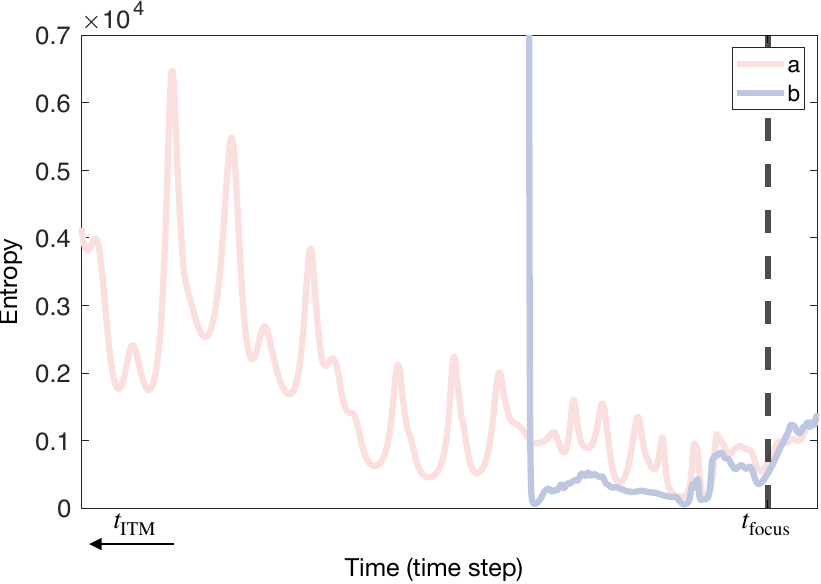}
    \caption{Entropy within the digital human phantom of the near surface scenario {\color{black}depicted in Fig.~\ref{fig:near}.} (a) {\color{black}Control:} homogeneous ITM. (b) {\color{black} Case study:} heterogeneous ITM {\color{black}with Steps I and II of the strategy applied. The temporal disruption is performed non-uniformly in air, as depicted by the white and dark regions in Fig.~\ref{fig:near}.} The steep drop in (b) is due to the sudden influx of reversed waves entering the phantom.}
    \label{fig:nearR}
\end{figure}

The resulting refocusing process is similar to what we have observed in parts (b) and (c) of the lossless study in Fig.~\ref{fig:simple}. As the source sits closer to the physical interface with air, the majority of the initial forward waves traveled into air quicker than in the deep seated source scenario. In fact, the homogeneous ITM in the control case was able to produce a focus with reasonable accuracy within $\Delta t_{\mathrm{focus}}$, resulting in a 26\% time persistence. However, due to attenuation in the biological tissues, the field maximum at $t_{\mathrm{focus}}$ falls on a neighboring trough near the skin, 17~mm away from the target. 

To minimize this effect, the experimental case is composed of a single wavefront selected with a 5\% threshold as described in Steps I and II but \emph{without} amplitude compensation as in Step III. Fig.~\ref{fig:near}b (left) shows the identified wavefront, which underwent a 100-fold $\epsilon_{\mathrm{r}}$ change. The result shows a time-reversed focus with 1~mm accuracy and a 55~mm$^2$ contour area at half maximum. The electric field distributions of cases (a) and (b) are similar but, in case (b), the strategy successfully weakens the problematic trough near the skin. As a result, the time persistence of case (b) improves to 100\%.

Shown in Fig.~\ref{fig:nearR} is the entropy of the control and experimental cases for the near surface source scenario, starting around the time when waves begin to enter the body for case (a). Notice they do not differ significantly near $t_\mathrm{focus}$, likely since the focus is primarily formed by anterior waves, with little contribution from the opposing direction.

By creating a reversed wave with a single crest and a single trough, this experiment shows that destructive interference is thereby minimized. For situations like this where a homogeneous ITM is sufficient to generate a reversed focus with reasonable accuracy and time persistence, applying a heterogeneous ITM using the wavefront selection methods alone without amplitude compensation is shown to suffice.

{\color{black}\section{Heterogeneous ITM on multi-target focusing}

The use of ITM for multi-target focusing in a homogeneous medium was demonstrated in~\cite{Crystal:2020}. The reversed waves are shown to refocus accurately on two separate targets, where the initial waves originated, provided the distance between them is at least a wavelength. This section investigates the robustness of utilizing heterogeneous ITM for multi-target focusing in a complex heterogeneous and lossy environment.

We review the multifocal scenario by placing both the near and deep seated sources in the bladder. The normalized initial wave field originating from the two sources is shown in Fig.~\ref{fig:twosrc}a (left), at the instant of the first speed disruption. For comparison, we again start with a control case using a homogeneous ITM. The result is comparable to the near surface scenario, as can be seen on the right in Figs.~\ref{fig:twosrc}a and~\ref{fig:near}a. This is to be expected since the amplitude of the waves reaching the surrounding air environment is much larger for those originating in the near surface source.

Our heterogeneous ITM approach for case (b) leverages the principle of superposition and combines the wavefront selection illustrated in Figs.~\ref{fig:deep}c and~\ref{fig:near}b, resulting in the selection in Fig.~\ref{fig:twosrc}b.
For the brown and white regions in Fig.~\ref{fig:twosrc}b, we used the same $\epsilon_{\mathrm{r}}$ change as in the deep seated case, i.e. 1.02-fold and 100-fold, respectively. Our experiments determined that a small 1.0005-fold $\epsilon_{\mathrm{r}}$ change in the blue region, which contains the large amplitude wavefront originating in the near surface source, generates two foci of comparable amplitude, shown on the right of the figure. In fact, the refocusing qualities of the two foci are similar to those in Section~\ref{sec:lossystudy}, when heterogeneous ITM was performed independently for each source.

In Fig.~\ref{fig:twosrcR}, we show the entropy for the two sources scenario, starting around the time when waves begin to enter the human body. As before, the successful convergence of the two foci in the experimental case (b) introduces a stronger local entropy minimum at $t_{\mathrm{focus}}$ when compared to the control case (a).

\begin{figure}[!t]
    \centering
    \includegraphics[width=\columnwidth]{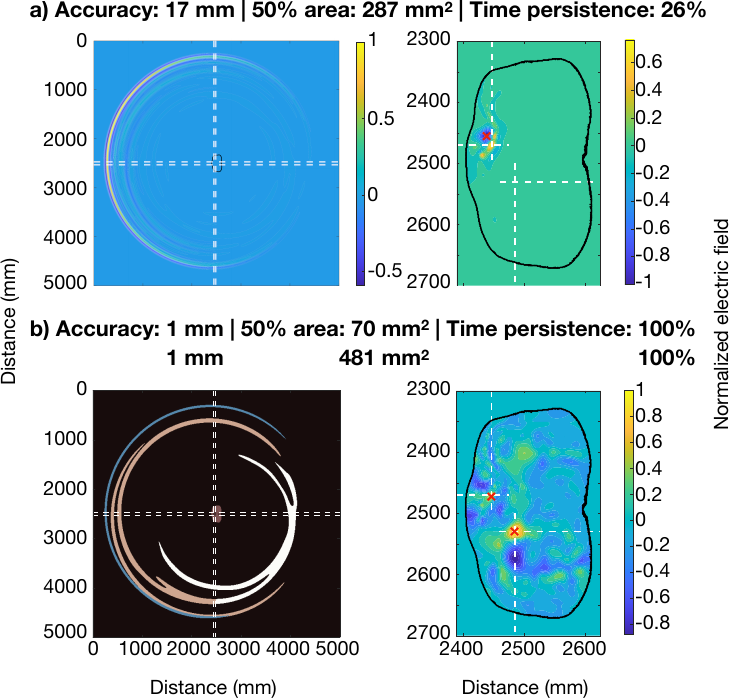}
    \caption{{\color{black}Combined deep seated and near surface scenario. (a) Control: homogeneous ITM. (b) Case study: heterogeneous ITM using the combined wavefronts of the single source cases shown in Figs.~\ref{fig:deep}c and~\ref{fig:near}b. On the left, illustration of (a) the initial wave field at the first disruption instant, and (b) the arrangement of the selected wavefronts for a three-strength ITM compensating the amplitude differences for the waves from the two sources.
    The results at $t_{\mathrm{focus}}$ are shown on the right.}}
    \label{fig:twosrc}
\end{figure}

\begin{figure}[!t]
    \centering
    \includegraphics[width=\columnwidth]{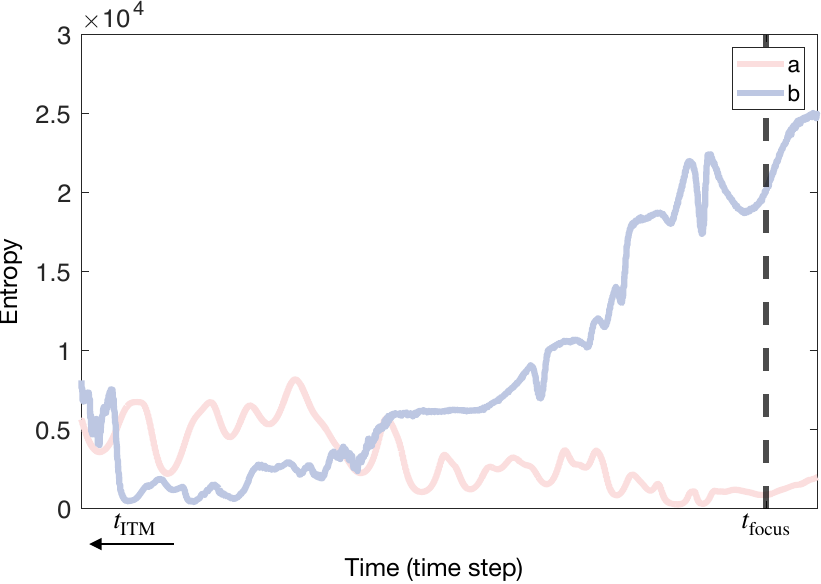}
    \caption{{\color{black}Entropy within the digital human phantom of the two source scenario depicted in Fig.~\ref{fig:twosrc}. (a) Control: homogeneous ITM. (b) Case study: heterogeneous ITM using the combined wavefronts of the single source cases shown in Figs.~\ref{fig:deep}c and~\ref{fig:near}b. The temporal disruption is performed non-uniformly in air, as depicted by the white, brown, blue, and dark regions in Fig.~\ref{fig:twosrc}.}}
    \label{fig:twosrcR}
\end{figure}
}

\section{Summary and Conclusion}

For electromagnetic waves, ITM can be conveniently achieved through a relative permittivity change. We first studied time-reversed foci in lossless environments and investigated the effects of only inducing ITM in selected regions of the space. The goal was to model reversed waves generated from partial initial waves and mimic regions with strong absorption elsewhere. The observations gathered in this initial study then guided the development of a technique to apply a {\em heterogeneous ITM} capable of compensating path dependent attenuation in lossy media.

To compensate the amplitude loss, a strategy is proposed which refines the implementation of ITM by using a single peak of the initial wave to reduce destructive interference at the focus and generating reversed waves with balanced intensity along opposing directions relative to the source. This is achieved by a non-uniform manipulation of the relative permittivity throughout space. 

We compare the time-reversed focus generated from homogeneous ITM and heterogeneous ITM in a 2D digital human phantom. In our first case study, a point source was placed deep inside the bladder. Although the results from homogeneous ITM showed a local peak at the target location, the field maximum was away from the target. The experimental results from heterogeneous ITM with amplitude compensation showed improved accuracy and time persistence. Our second case study, where the source is closer to the physical interface between the human body and air, shows heterogeneous ITM without amplitude compensation may suffice to shift the focus to the target in cases where careful selection of the wavefronts is enough to improve accuracy and time persistence. {\color{black}Additionally, we examined a multifocal case using two sources which demonstrated that heterogeneous ITM can also compensate the amplitude differences seen in waves originating from different sources by simply leveraging superposition.}

In this paper, heterogeneous ITM is introduced as a means to adjust the amplitude of the produced reversed waves by considering the geometry of the problem and the electric field amplitude of the initial forward waves. Although the strategy is shown to be promising, we acknowledge its practical feasibility might be challenging due to the necessary knowledge of the wave configuration at the time of ITM and the ability to alter the relative permittivity with high spatio-temporal precision. As part of ongoing work, a complete protocol is being developed for the precise identification of both the targeted wavefronts and the extent of the ITM disruptions to maximize the benefits of this procedure.

\section*{Acknowledgment}

The digital human phantom, based on MRI and provided by RIKEN, Japan, is approved for use by their ethical committee under a non-disclosure agreement with The University of Manchester, UK. The biological data were fitted to the one-pole Debye model by José González and Aleksandr Kudasev.

\ifCLASSOPTIONcaptionsoff
  \newpage
\fi



\bibliographystyle{IEEEtran}
\bibliography{bibAll.bib}{}
%\begin{thebibliography}{1}


%\end{thebibliography}


\vspace*{-6.6mm}
\begin{IEEEbiography}[{\includegraphics[width=1in,height=1.25in,clip,keepaspectratio]{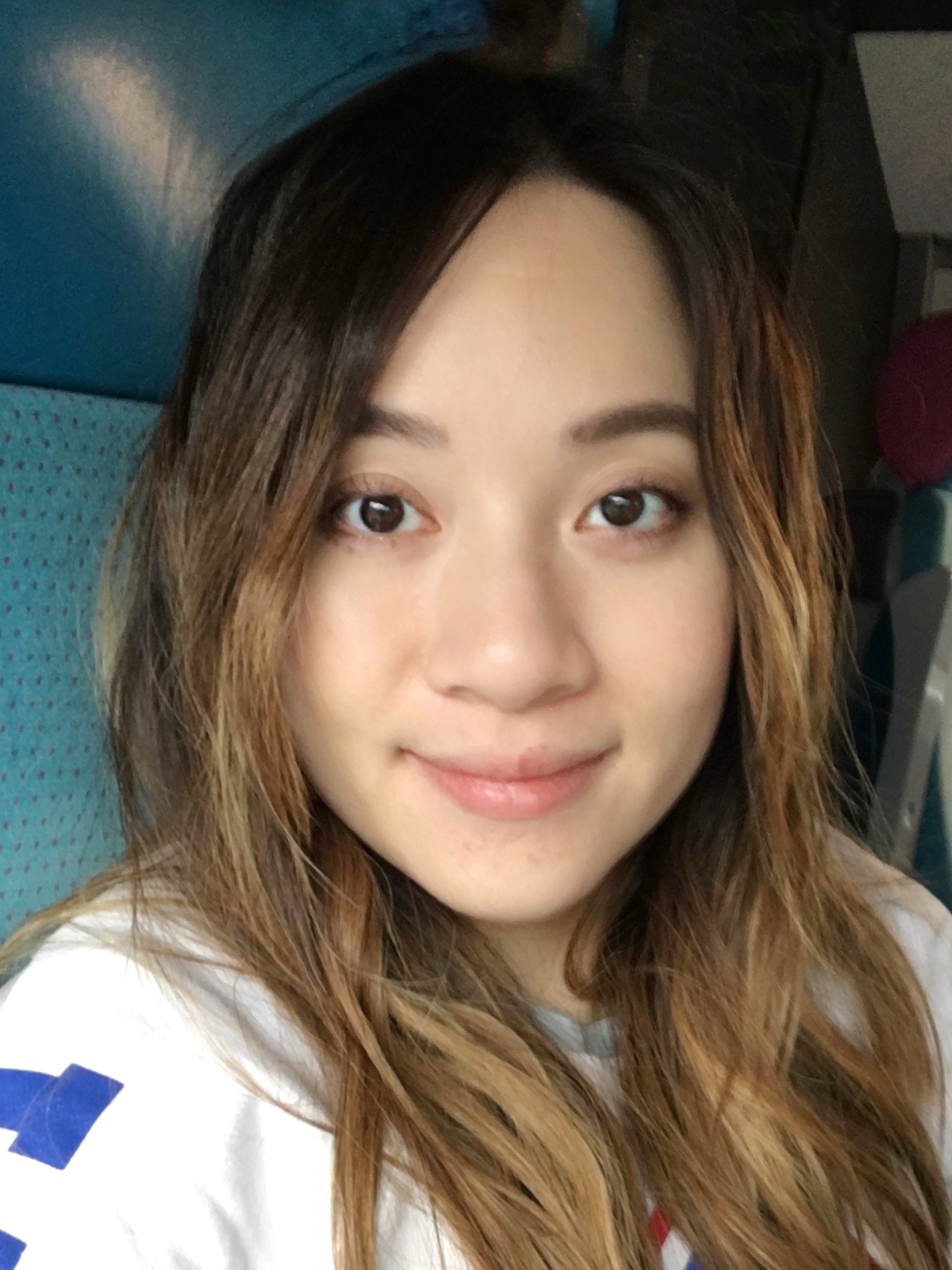}}]{Crystal T. Wu}
received the B.S. degree in Biochemistry and Molecular Biology from Michigan State University and the M.Sc. degree in Computer Science from The University of Manchester, where she is currently pursuing the Ph.D. degree in the Department of Computer Science, and in collaboration with the Department of Electrical and Electronic Engineering since 2018. Her research interests include instantaneous time mirrors, computational electromagnetics, and attenuation compensation techniques in wave propagation.
\end{IEEEbiography}

\begin{IEEEbiography}[{\includegraphics[width=1in,height=1.25in,clip,keepaspectratio]{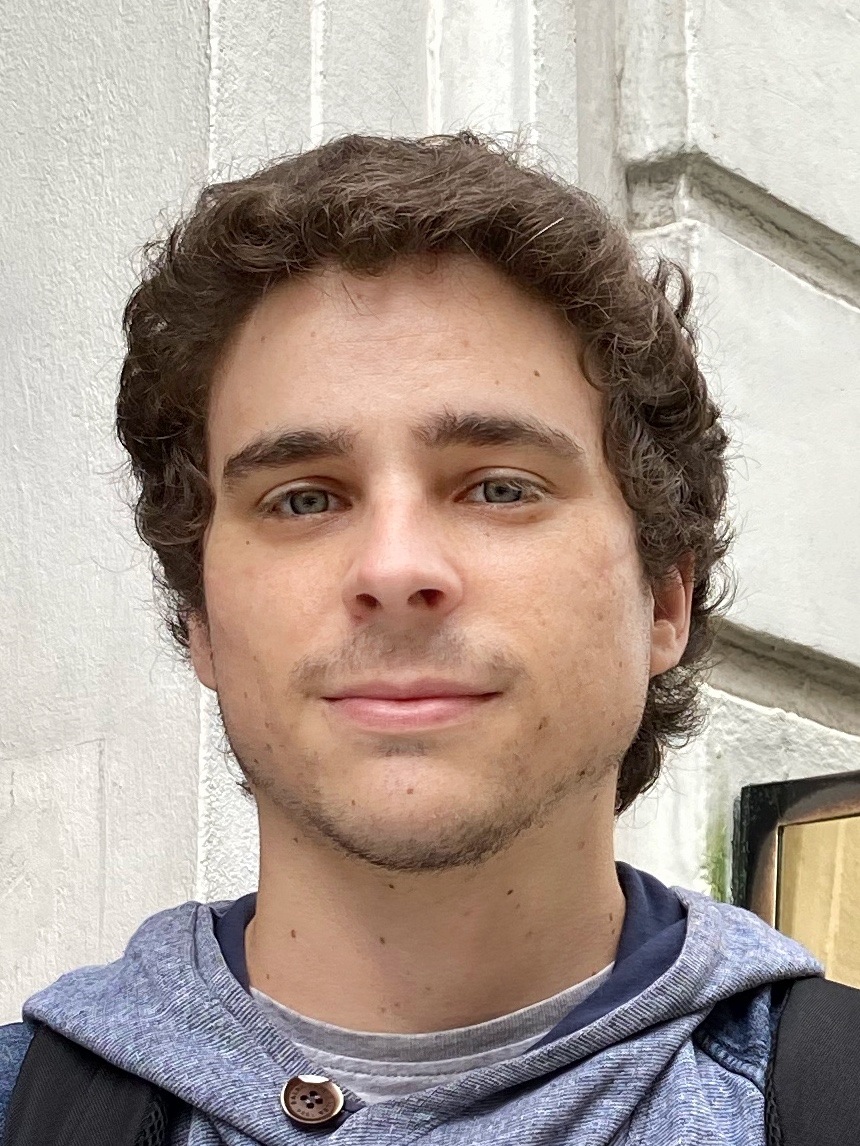}}]{Nuno M. Nobre}
holds an M.Sc. degree in Engineering Physics from the University of Coimbra and a Ph.D. degree in Computer Science from the University of Manchester. His research interests include numerical modeling and techniques for automatic optimization of scientific programs with portability and scalability potential across computing platforms.
\end{IEEEbiography}

%\newpage

\begin{IEEEbiography}[{\includegraphics[width=1in,height=1.25in,clip,keepaspectratio]{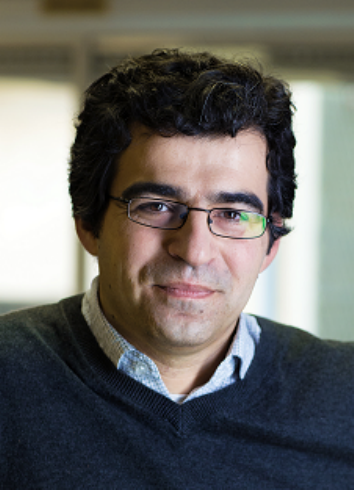}}]{Emmanuel Fort\textnormal{,}}
full professor at ESPCI Paris, PSL University, holds the AXA Chair in Biomedical Imaging at the Langevin Institute. He is internationally recognized in the field of fluorescence imaging and plasmonic sensing for biological and medical applications. He has developed several innovative super-resolution techniques and nanostructured materials for applications in microarray sensors. He is also a pioneer in hydrodynamic quantum analog notably by introducing self-propelled bouncing droplets on the surface of a vibrated liquid as the first classical wave-particle duality. He has also made major contributions to the field of time varying media to control wave propagation with the introduction of instantaneous time mirrors and to dynamically stabilize liquid interfaces. He received the Denis Diderot Innovation Award and the Jerphagnon Prize. He is co-inventor of 8 patents and co-founder of the startup Abbelight. He has published more than 70 articles in internationally renowned journals.
\end{IEEEbiography}

\begin{IEEEbiography}[{\includegraphics[width=1in,height=1.25in,clip,keepaspectratio]{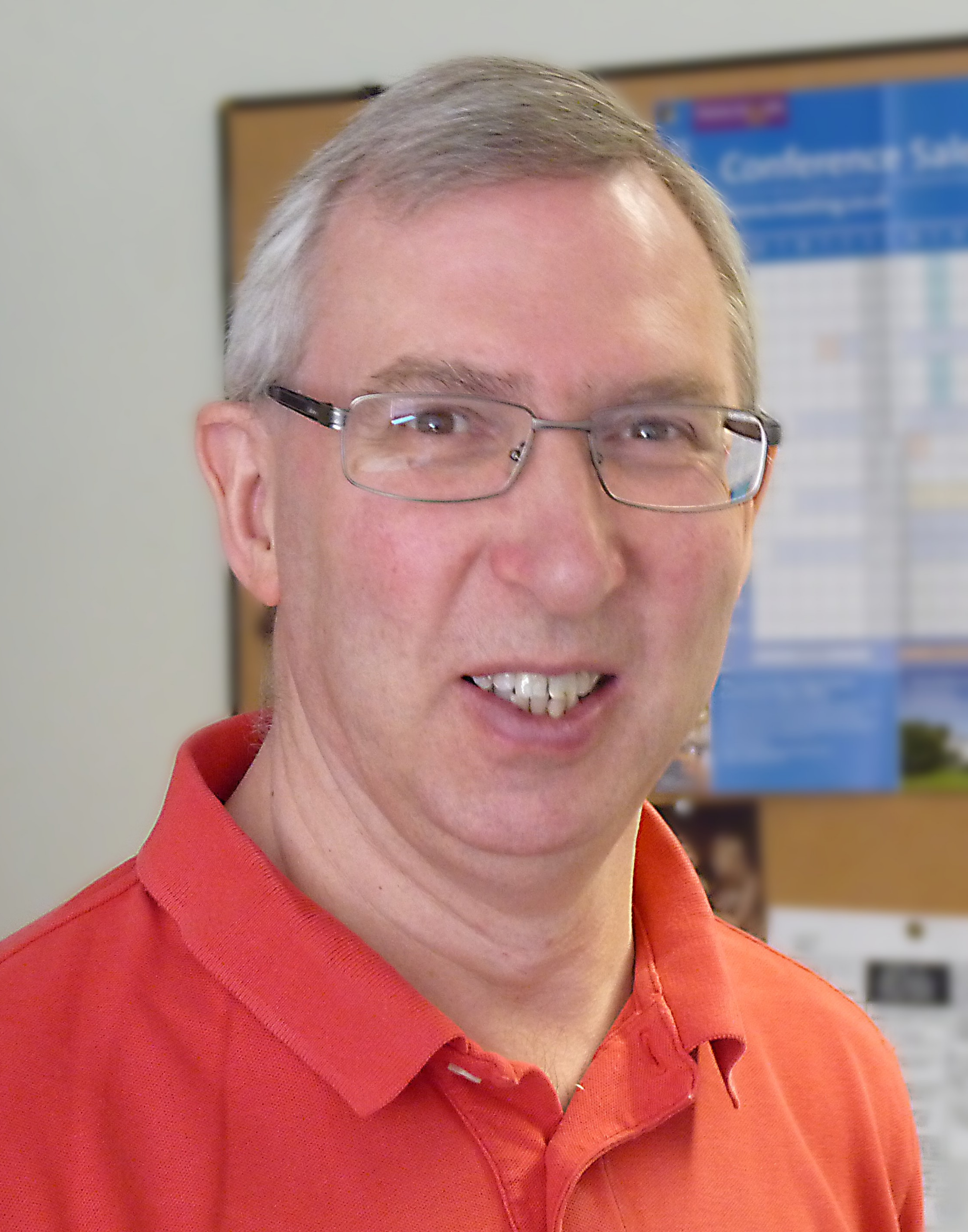}}]{Graham D. Riley}
is a senior lecturer in Computer Science with the Advanced Processor Technologies Group at the University of Manchester. His research interests include methods and techniques for performance analysis and performance improvement for scientific and engineering/physics applications, including climate modelling and machine learning, on heterogeneous parallel platforms from desktop to extreme-scale.
\end{IEEEbiography}

\begin{IEEEbiography}[{\includegraphics[width=1in,height=1.25in,clip,keepaspectratio]{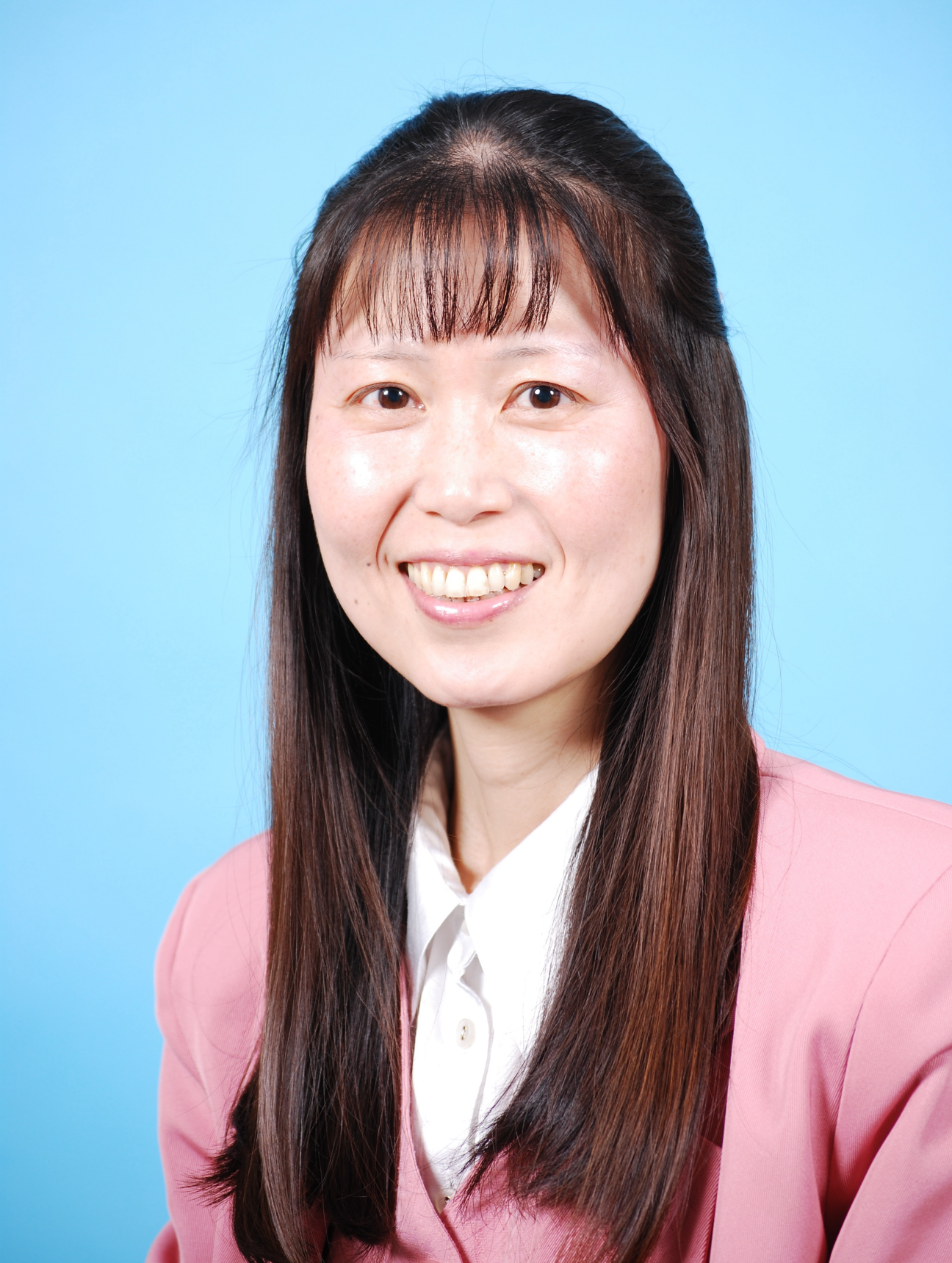}}]{Fumie Costen}
(M'07-SM'12) received the B.Sc. degree, the M.Sc. degree in electrical engineering and the Ph.D. degree in Informatics, all from Kyoto University, Japan.

From 1993 to 1997 she was with Advanced Telecommunication Research International, Kyoto, where she was engaged in research on direction-of-arrival estimation based on Multiple Signal Classification algorithm for 3-D laser microvision. She filed three patents from the research in 1999 in Japan. She was invited to give 5 talks in Sweden and Japan during 1996-2014. From 1998 to 2000, she was with Manchester Computing in the University of Manchester, U.K., where she was engaged in research on metacomputing and has been a Lecturer since 2000. Her research interests include computational electromagnetics in such topics as a variety of the finite difference time domain methods for microwave
frequency range, microwave imaging, application of deep learning to FDTD, radar images and medical images. She filed a patent from the research on the boundary conditions in 2012 in the U.S.A. Her work
extends to the hardware acceleration of the computation
using general-purpose computing on graphics processing units, Streaming Single Instruction Multiple Data Extension and Advanced Vector eXtentions instructions.

Dr. Costen received an ATR Excellence in Research Award in 1996 and a best paper award from 8th International Conference on High Performance Computing and Networking Europe in 2000 and Teaching Excellence award 2021.
\end{IEEEbiography}

%\vfill


\end{document}